\documentclass[usenatbib,useAMS]{mn2e}
\usepackage{graphicx}
\usepackage{times}
\usepackage{aas_macros}
\title[]{A $J$-band detection of the sub-stellar mass donor in SDSS J1433+1011}

\author[]{S.\,P.\ Littlefair$^{1}$, C.\,D.\,J.\, Savoury$^{1}$, V.\,S.\, Dhillon$^{1}$, T.\,R.\, Marsh$^{2}$, B.\,T.\, G\"{a}nsicke$^{2}$, \newauthor T. Butterley${^3}$, R.\, W. Wilson${^3}$, J. Southworth$^{4}$ \& C.\,A.\, Watson$^{5}$ \\
$^1$Dept of Physics and Astronomy, University of Sheffield, Sheffield, S3 7RH, UK \\
$^2$Dept of Physics, University of Warwick, Coventry, CV4 7AL, UK\\
$^3$Centre for Advanced Instrumentation, University of Durham, South Road, Durham, DH1 3LE, UK\\
$^4$Astrophysics Group, Keele University, Staffordshire, ST5 5BG, UK\\
$^5$Astrophysics Research Centre, QueenÕs University Belfast, Belfast BT7 1NN, UK\\}

\date{\center{\Large Submitted for publication in the Monthly Notices of the
Royal Astronomical Society \\ 
\vspace{.5cm} \today}} 

\begin{document}
\maketitle

\begin{abstract} 
We present time-resolved $J$-band spectroscopy of the short period cataclysmic variable SDSS~J143317.78+101123.3. We detect absorption lines from the sub-stellar donor star in this system, which contributes $38\pm5$\,\% to the $J$-band light. From the relative strengths of the absorption lines in the J-band, we estimate the spectral type of the donor star to be L$2\pm1$. These data are the first spectroscopic detection of a confirmed sub-stellar donor in a cataclysmic variable, and the spectral type is consistent with that expected from semi-empirical evolutionary models.

Using skew mapping, we have been able to derive an estimate for the radial velocity of the donor of ${\rm K_d} = 520\pm60$ km\,s$^{-1}$.  This value is consistent with, though much less precise than, predictions from mass determinations found via photometric fitting of the eclipse light curves. 
\end{abstract} 


\section{Introduction}
\label{sec:introduction}
Cataclysmic variables (CVs) are semi-detached binary stars consisting of a white dwarf primary and a Roche-lobe filling donor star. One of the key features of the orbital period distribution of CVs is the sharp cut-off at orbital periods of $\approx$80 minutes - the so-called period minimum. The location of the period minimum is closely linked to the thermal equilibrium state of the donor star. At long periods, the donor shrinks in response to mass loss, causing the CV to evolve towards shorter orbital periods. As the CV evolves, the donor's thermal timescale increases faster than the mass-loss timescale. As a result, the donor is pushed further and further from thermal equilibrium and eventually reaches a point where the radius no longer shrinks, and may even expand, in response to mass loss. As a result, the binary begins to evolve towards longer orbital periods \cite[see e.g.][for more details]{knigge11a}. Systems which have evolved past the period minimum are known as period-bouncers.

It has long been recognised that there may be problems with this picture. Firstly, theoretical predictions of the period minimum ($P_{min}\approx65$ minutes) are substantially shorter than the observed value of $P_{min} \approx 82$ minutes \citep{gaensicke09}. Secondly, evolutionary models predict that the CV population should be dominated by period-bouncers \cite[e.g.][]{kolb93}. Despite this, for a long time there was no direct evidence for the existence of period bounce CVs \citep{littlefair03}. The difficulty in establishing the contribution of period-bounce CVs to the population as a whole arises as a result of the faintness of the donor, relative to the accretion disc and white dwarf in these systems. This difficulty was side-stepped by measuring donor masses using a photometric eclipse-fitting method, yielding the first unambiguous period-bouncers \citep{littlefair06}.  Applying this technique to a small sample of 14 CVs, \cite{savoury11} estimate that $\sim$15\% of all CVs have evolved past the period minimum. Unfortunately, due to the small number of short-period eclipsing CVs, improving the precision of this estimate relies on establishing alternative methods of identifying period-bounce CVs.

One potential method for identifying period bouncers is to look for direct evidence of the donor star in the near infrared. Several authors have presented evidence for donor stars with spectral types later than M \cite[e.g.][]{mennickent04,unda-sanzana08,aviles10}, however it is not easy to identify these systems as period bouncers, since the spectral type at which period bounce occurs is not well known. Recently, \citealt{knigge11b} (hereafter K11) produced a semi-empirical evolutionary track for CVs, by taking state-of-the-art  stellar models and tuning the angular momentum loss rates so that the model predictions fit mass-radius measurements of CV donors. K11 predict that the period minimum occurs at a period of $P_{min}=81.8$ minutes, in excellent agreement with the observed value. In their models, this corresponds to a donor mass of $M_d = 0.061 M_{\odot}$ and a spectral type of M9.5. Thus, if the evolutionary model of K11 is accurate, donors with an L-type spectral class are likely to be period bouncers. The caveat to this conclusion is that the spectral type-orbital period relationship of K11 is largely unconstrained by observations at short orbital periods. Indeed, there are no examples in the literature of a robust period-bounce CV with an accurate measurement of the donor star's effective temperature.  In this paper we present time-resolved $J$-band spectroscopy of SDSS~J143317.78+101123.3  (hereafter SDSS J1433), with the aim of detecting the substellar donor star.  SDSS J1433 has a measured donor star mass of $M_d = 0.0571\pm0.0007$\,M$_{\odot}$ \citep{savoury11} and an orbital period of 78.11 minutes  \citep{littlefair08}, making it a strong candidate for a period bounce CV. 

\section{Observations}
\label{sec:obs}
SDSS J1433 was observed using the near-infrared spectrograph NIRI \citep{hodapp03} on Gemini North, Mauna Kea, Hawaii. Observations were taken in service mode. NIRI was configured with the f/6 camera, Aladdin In-Sb 1024$\times$1024 pixel array, a 0.75$\times$110 arc-second slit and $J$-grism disperser with a dispersion of $\sim$0.133\AA\,pixel$^{-1}$ at 12500\AA. The resolving power, as measured from the width of night sky lines, was found to be $R=454\pm16$.

The data were obtained on nights beginning 2 July 2009, 7 May 2010, 26 May 2010, 27 May 2010, 15 June 2010 and 16 June 2010. In total 389 spectra covering a wavelength range of 10354--13959\,\AA\, were obtained. Exposure times were 59 seconds. The data were taken using a standard nodding pattern, moving the target along the slit by a few arc-seconds between exposures to improve sky subtraction. Observations of the G2V standard star HIP 73593 were taken at the start of each night and again after every hour of science data. Exposure times for the standard star were 3 seconds. The standard star observations were used both to flux calibrate the spectra and to correct for telluric absorption. Arc spectra were taken after every hour of science data using an argon lamp. Seeing conditions on each night were good, with typical seeing between 0.4 and 1.2 arc-seconds. 

\section{Data Reduction}
\label{sec:red}
The NIRI images were initially processed using the Gemini IRAF package. Quartz-halogen flat frames were taken with the Gemini Facility Calibration Unit (GCAL). A set of nine flat frames were taken for each hour of science data. The  flat field was constructed by taking the mean of the individual frames. Spatial structure was removed, and the flat field was normalised, by dividing by a spline fit to the flat field, after collapsing in the spectral direction.  During this process, a bad pixel map was produced by flagging normalised pixels that fall above or below values of 1.2 and 0.8 respectively.

Frames taken at adjacent nodding positions were subtracted from each other; as well as providing a first pass at sky subtraction, this removes bias and dark current signals from the data. Each complete batch of NIRI observations were taken in groups of 54 spectra. These 54 spectra were combined in groups of four, to improve the signal-to-noise ratio of the data, without significantly degrading the time-resolution. The frames were aligned using the World Coordinate System (WCS) information stored in the fits headers of each file. Individual frames were combined using a clipped mean, with outlying points rejected at the 5$\sigma$ level.

Spectra were extracted using the {\sc pamela} data reduction package\footnote{http://deneb.astro.warwick.ac.uk/phsaap/software/pamela/html/INDEX.html}. The location of the spectrum was tracked across the CCD, then this location was fit using a second-order spline function. Object and sky regions were defined by hand and then the data were extracted using an optimal extraction technique as described by \cite{marsh89}. A second sky subtraction was performed by subtracting a second-order polynomial fit to the background regions. Wavelength calibration was performed using a fourth-order polynomial fit to the position of argon emission lines in the arc spectra. RMS residuals to this fit were of the order of 0.3\AA.

Observations of the G2V standard star  were used to flux calibrate and telluric correct the target spectra as follows. The standard star spectrum was normalised using a seventh-order polynomial fit. A telluric corrected Solar spectrum, taken at Kitt Peak Observatory and produced by NSF/NOAO, was broadened to match the resolution of the NIRI data and normalised using a seventh-order polynomial fit. The Solar spectrum was scaled to match the strength of absorption lines in the standard star spectrum, and subtracted from the standard star, leaving only the telluric features. This telluric spectrum was multiplied by the polynomial fit to reproduce the original standard star spectrum, minus any stellar absorption lines.  The resulting spectrum was divided by a black body spectrum of 5778K, scaled to match the reference flux of the standard star \citep{cutri03,tokunaga05}. The individual target spectra were divided by the result to provide flux calibrated, telluric corrected data. 

\section{Results}
\subsection{Average spectrum}
\begin{figure}
\begin{center}
\includegraphics[scale=0.45,clip=true,trim=80 50 0 0]{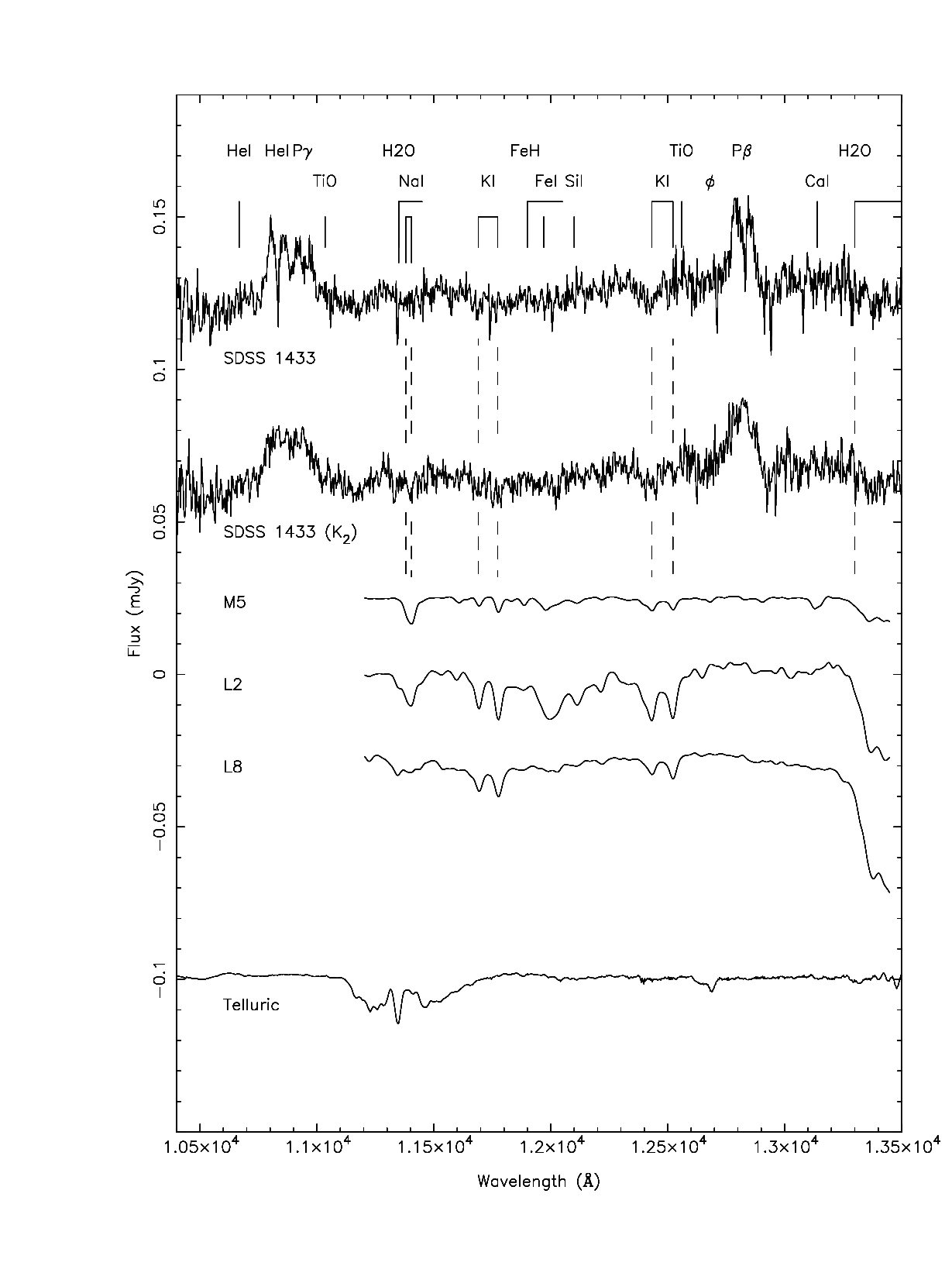}
\caption{The average $J$-band spectrum of SDSS J1433. The upper spectrum of SDSS J1433 is in the rest frame of the binary, and is offset by 0.06 mJy from the absolute flux. The lower spectrum of SDSS J1433 has been corrected for the predicted radial velocity of the donor \protect\citep{savoury11} and thus represents an average in the rest frame of the donor. Also shown are IRTF  library spectra of M5, L2 and L8 templates and the telluric spectrum. The template and telluric spectra have been normalised and offset for clarity.}
\label{fig:average}
\end{center}
\end{figure}
Figure~\ref{fig:average} shows the average $J$-band spectrum of SDSS J1433 in the rest frame of the binary and the rest frame of the donor star. Template stars from the IRTF spectral library \citep{cushing05} are also shown, after having been broadened to account for the resolution of the NIRI data and the predicted rotational broadening of the donor star \citep{savoury11}. The average $J$-band flux (across all orbital phases) is $0.064\pm0.004$ mJy, giving an apparent $J$-band magnitude of ${\rm J} = 18.5 \pm 0.1$. Note that this value is not corrected for differential slit losses between the target and standard star.

The average spectrum of SDSS J1433  shows increasing flux towards the red end of 
the spectrum. The slope of the continuum was measured at ($3.30 \pm 0.13$)$\times$10$^{-2}$ 
mJy/$\mu$m. The presence of a red continuum could reasonably be attributed to the donor 
star because the white dwarf and accretion disc are expected to have a blue continuum 
\citep[see, e.g. Fig 3.2 of][]{hellier01}. The presence of a red continuum alone is insufficient evidence 
for a detection of the secondary star, since cyclotron emission may also contribute, and flat-fielding and flux-calibration issues can also affect the continuum slope.

There is evidence of a donor star absorption feature around 13300~\AA~that is attributed to 
the headless water band at 13300~\AA. There is also some tentative evidence of K I absorption 
around 12432~\AA. The 0-1 FeH band at 12000~\AA\, is not visible, though the broad nature of this feature means it may be masked by noise (see figure~\ref{fig:optsub_fit}, for example). The structure in the CV spectrum around 13000~\AA\, may be due to inadequate telluric correction, or to genuine features in the CV spectrum. We believe the latter explanation is more likely, as there is no evidence for the sharp telluric feature at 11350~\AA, the peaks and troughs of the telluric spectrum do not line up well with the troughs and peaks of the CV spectrum, and the continuum shape of the CV spectrum appears to follow a donor star template well, even inside the telluric affected region (see figure~\ref{fig:optsub_fit}). When the spectrum is corrected for the predicted orbital motion of the donor star  \citep{savoury11}, the second component of the potential K I absorption doublet around 12477~\AA~becomes visible and the water band, which was previously visible,  appears sharper. The improvement in the water band adds confidence that this is not an artefact of improper telluric calibration. There is no evidence of H$_{2}$O or CH$_{4}$ absorption around 11000~\AA~in either spectra, which, for mid to late T-dwarfs, should be similar in strength to that of the water band at 13300~\AA\ \citep[e.g.][]{kirkpatrick05}.

\subsection{Skew Mapping}
\label{sec:skew_mapping}
Given the faintness of the potential donor star features and the poor signal-to-noise ratio of the data, 
there is an element of uncertainty as to whether the features described above are genuine absorption features  or merely the product of  noise in the spectrum. However, if these features are real, then a {\em skew map} should produce a clear peak, centred on the radial velocity of the donor star, $K_d$. Skew mapping is a tomographic technique used to measure the radial velocity of the donor star when spectral features are too weak for conventional cross-correlation techniques \citep{smith93}.

\begin{figure}
\begin{center}
\includegraphics[scale=0.4,angle=270]{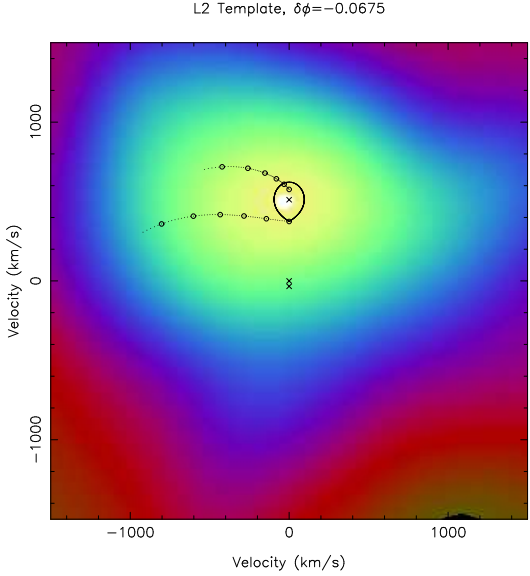}
\caption{Skew map for SDSS 1433 produced using an L2 template. The predicted position of the donor star and the path of the gas stream are marked. The crosses on the map are, from top to bottom, the centre of mass of the donor star, the centre of mass of the system, and the centre of mass of the white dwarf. These crosses, the Roche lobe of the donor,  the path of the gas stream (lower curve), the Keplerian velocity at locations along the gas stream (upper curve) are plotted using the values of mass ratio, donor velocity and white dwarf velocity estimated by \protect\cite{savoury11}.}
\label{fig:sdss1433_skew_map}
\end{center}
\end{figure}

We produced skew maps from our data as follows. The data were phase binned into 15 bins according to the ephemeris of \cite{littlefair08} and a range of template spectra from M5 to L8 were obtained from the IRTF spectral library \citep{cushing05}. The template spectra were broadened to match the predicted rotational velocity of the donor star of $v \sin i = 99.7$ km s$^{-1}$ \citep{savoury11}. The amount of broadening required for each template was calculated by obtaining rotational velocities for our templates from the literature \citep{tinney98, mohanty03, reiners07, reiners08, morin10, deshpande12} and adding an additional amount of broadening in quadrature to match the total rotational broadening above. The templates were further broadened to match the resolution of the NIRI data. The templates and phase-binned CV spectra were then normalised by dividing by a first order polynomial. This ensured that relative line strengths across the CV spectrum was preserved. Both the CV and template spectra were binned onto the same wavelength scale (11203-13450~\AA~in 624 pixels). The P$\beta$ emission line was masked to avoid accretion features contaminating the results.  Skew maps were then produced for SDSS J1433 by cross-correlating the phase-binned spectra of SDSS J1433 with the template stars, and then back-projecting the cross-correlation functions as described in \cite{smith98}. We assumed a systematic velocity of the binary of $\gamma=0$ km\,s$^{-1}$. Figure \ref{fig:sdss1433_skew_map} shows the skew map of SDSS J1433 made with the L2 template,  found in section~\ref{sec:spectraltype} to be a reasonable estimate of the spectral type of the donor star.

As stated above, if the donor star absorption features are real, we would expect a peak at ($K_x$, $K_y$) = (0, $K_d$) in the skew map. When we first performed the skew mapping a strong peak was visible, but significantly offset from the $K_{x}$ = 0 axis. One explanation for this offset is a discrepancy between the ephemeris of \citet{littlefair08} and the ephemeris at the time of our observations. To check this, follow-up photometry of the eclipse of SDSS J1433 was obtained in March 2012 using a robotic 50cm telescope on La Palma\footnote{http://sites.google.com/site/point5metre}. These observations revealed a phase lag $\Delta\phi = +0.09$ between the observed mid-eclipse time and the mid-eclipse time predicted by the ephemeris of \citet{littlefair08}.  This difference is not consistent with the statistical errors on the ephemeris. The rate of change in orbital period (${\dot{P}}$) required to create a phase shift of $\Delta\phi = +0.09$ is $(5.2\pm0.5) \times 10^{-8}$ days yr$^{-1}$, which is several orders of magnitude larger than expected for systems near the period minimum (K11). Further monitoring of the eclipses of SDSS J1433 is highly desirable to determine if this rapid period change is compatible with the presence of a third body. Given the results above, it is highly likely that the ephemeris used to phase bin the data may not be valid for the spectroscopic observations obtained in 2009 and 2010, and that a phase offset may be appropriate. A phase offset of $\Delta\phi = -0.0675$ was found to be sufficient to shift the peak of the skew map onto or close to the $K_{x}$ = 0 axis, and the skew map shown in figure~\ref{fig:sdss1433_skew_map} has had this offset applied.

The skew map shown in figure~\ref{fig:sdss1433_skew_map} shows a significant peak at $K_{x}$ = $-30 \pm 50$ km s$^{-1}$,  $K_{y}$ = $520 \pm 60$ km s$^{-1}$. The uncertainties were derived using a bootstrapping technique, where the cross-correlation procedure was repeated, each time with one different phase bin masked. The uncertainties from this technique, $\sigma_{K_{x}}$ and $\sigma_{K_{y}}$, were found to be $\sigma_{K_{x}} = 30$ km s$^{-1}$ and $\sigma_{K_{y}}$ = 40 km s$^{-1}$. The change in position of the peak was then measured after cross-correlating against a range of different template stars, to find the error introduced by the uncertain spectral type. This gave $\sigma_{K_{x}}$ = 40 km s$^{-1}$ and $\sigma_{K_{y}}$ = 50 km s$^{-1}$. These values were then added in quadrature to give the final uncertainty. The skew map thus provides a measure for the donor star's radial velocity of $K_d = 520\pm60$ km s$^{-1}$. This value is in good agreement with the radial velocity predicted by \cite{savoury11} of $K_{d}$ = $511 \pm 1$ km s$^{-1}$, providing strong evidence that the absorption features seen in figure~\ref{fig:average} genuinely arise from the donor star. In addition, it provides further evidence to support previous claims that the photometric technique used to derive component masses  in eclipsing CVs is robust \citep{savoury12, copperwheat10, tulloch09}.

\subsection{The spectral type of the donor star}
\label{sec:spectraltype}
The spectral type of the donor star was estimated by comparing the relative strengths of absorption lines in the $J$-band with those of template spectra via an optimal subtraction technique as follows. A donor-star rest-frame average spectrum of SDSS J1433 and template spectra were prepared as described in section~\ref{sec:skew_mapping}. We then subtracted a scaled version of the template spectrum from the spectrum of SDSS J1433, with the scaling chosen to minimise $\chi^2$ between the residual spectrum and a smoothed version of the residual (a Gaussian with FWHM=70 pixels was used for this smoothing). By repeating this procedure with templates of different spectral type, the correct spectral type can be identified as the one with the lowest $\chi^2$, and the value of the scaling constant provides the fractional contribution of the donor to the $J$-band light.

\begin{figure}
\begin{center}
\includegraphics[scale=0.3,angle=270]{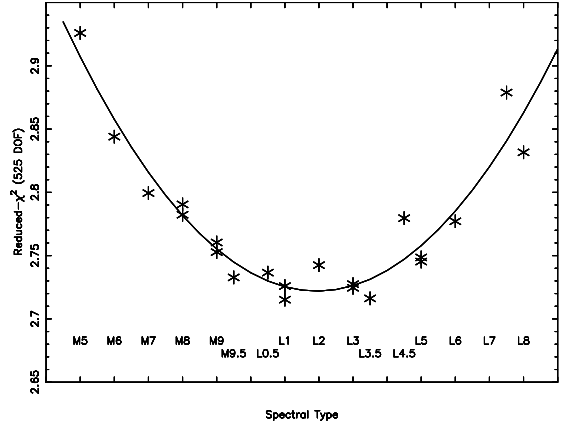}
\caption{Reduced-$\chi^{2}$ versus spectral type from the optimal subtraction technique. See section~\protect\ref{sec:spectraltype} for details.}
\label{fig:sdss1433_optsubs}
\end{center}
\end{figure}

\begin{figure}
\begin{center}
\includegraphics[scale=0.35,angle=270]{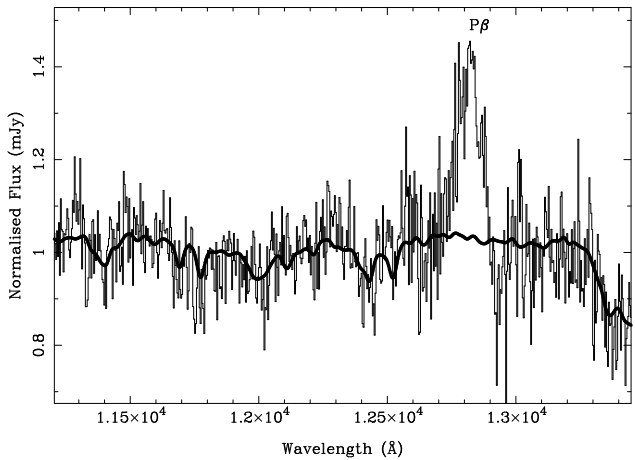}
\caption{The normalised, donor-star rest-frame average spectrum of SDSS J1433 with the best fitting template spectrum (Kelu1AB - L2V) over plotted.  The template has been prepared as described in section~\ref{sec:skew_mapping},  scaled by a factor of 0.38 and has had a 2nd order polynomial added, so that the continuum slope matches that of the CV spectrum.}
\label{fig:optsub_fit}
\end{center}
\end{figure}

Figure \ref{fig:sdss1433_optsubs} shows the resulting plot of $\chi^{2}$ versus spectral type. A third order polynomial fit to $\chi^2$ versus spectral type indicates a best fitting spectral type of L2. The best spectral type was found to vary slightly  depending on the level of smoothing used. In order to estimate the uncertainty in spectral type, the optimal subtraction routine was repeated with different levels of smoothing ($15-200$ km s$^{-1}$) and the resulting change in the best spectral type measured. The best-fitting spectral type was found to vary between L1.5 and L3, and so the uncertainty in spectral type was estimated to be $\pm$1 spectral type class. Including the uncertainty in the spectral type, the percentage contribution of the donor star to the $J$-band light is found to be $38\pm5$\,\%. Figure~\ref{fig:optsub_fit} shows the best fitting template spectrum scaled and over-plotted on the average CV spectrum.

\subsubsection{Caveats}
Although the relative strength of the absorption lines in the $J$-band spectrum indicate a best-fitting spectral type of L2, this result needs to be interpreted with caution. This is because the infrared spectral appearance of L and T dwarfs is not a simple proxy for effective temperature, as is the case with warmer objects. The infrared spectral region is strongly influenced by opacity from collision-induced H$_2$ absorption and dust grains. These opacity sources are, in turn, strongly influenced by metallicity and gravity, meaning the infrared spectral type of L and T-dwarfs depends on multiple parameters \cite[e.g][]{kirkpatrick08}.

This complication reveals itself in multiple ways. The first is that the effective temperatures of objects at a single spectral type can show significant scatter of a few hundred Kelvin \citep{kirkpatrick05}. The second is that objects with identical spectral types in the optical can show markedly different infrared spectra \cite[see figure 3 of ][for example]{kirkpatrick08} - in our data this is probably the reason for the increased scatter in figure~\ref{fig:sdss1433_optsubs} at later spectral types. Lastly, the sensitivity of the infrared regions to multiple parameters means that spectral types and effective temperatures derived from different infrared spectral bands can differ from each other by as much as 700K \citep{cushing08}. As a result of this last factor, the detection of the donor star in the $J$-band light of SDSS 1433 makes spectroscopy covering a wide range of wavelengths highly desirable in future. 

\section{Discussion}
\label{sec:discussion} 
For the substellar donor in SDSS J1433, the expected effective temperature will depend upon the mass-loss and thermal history of the donor. An indication of the expected spectral type can be found via comparison to the semi-empirical evolutionary sequence of K11. However, this comparison is complicated by a number of factors. First, the orbital period of SDSS J1433 (78.11 minutes  - \citealt{littlefair08}) is {\em shorter} than the minimum period of the semi-empirical evolutionary track in K11 (82 minutes). Second, the white dwarf mass in SDSS J1433 is 0.87M$_{\odot}$, as compared to the canonical mass of 0.75M$_{\odot}$ used by in K11. With all other parameters held constant, the higher white dwarf mass would lead to an increased angular momentum loss rate at a given period. This would result in an increased  minimum period and increased   effective temperature, at a constant orbital period \cite[see figure 3 of][for example]{kolb99}. However, from inspection of the population synthesis models for CVs  \citep{howell01} we judge that the effect of the increased white dwarf mass on the evolution of SDSS J1433 is small.

Instead of comparing our spectral type directly with the predictions of K11, we chose convert our spectral type to an effective temperature, and compare with the effective temperature predicted by K11 at the donor mass of SDSSJ1433 (0.057 $\pm$ 0.007\,M$_{\odot}$) . This is because whilst K11 derive the donor star's effective temperature directly from the \cite{baraffe98} stellar models, the spectral types are obtained via a less direct route. The effective temperatures are used to produce colours from atmospheric model grids, and a colour-spectral type relationship is used to assign a spectral type. We convert our spectral type of L2$\pm$1 for the donor in SDSS J1433 to an effective temperature using the empirical relationship between infrared spectral type and effective temperature of \cite{stephens09}. Taking into account the observed $\sim$100\,K scatter around this empirical relationship, this yields an effective temperature of 2000$\pm$200\,K. This is in remarkable agreement with the predicted value of 2000$\pm$40\,K from K11.

The orbital period-spectral type relationship of K11 is an extremely steep function of orbital period for objects around period bounce. For a small change in orbital period from 82.3 to 82.7 minutes, the spectral type of the donor evolves from L0 to L4! As a result, the donor star spectral type can be a powerful diagnostic of the evolutionary state of the CV. K11 find that the period minimum occurs at a donor mass of $M_d = 0.061 M_{\odot}$, for a wide range of mass transfer rates. Hence, with a donor mass of $M_d = 0.0571\pm0.0007$ \citep{savoury11}, it is extremely likely that SDSS J1433 has evolved past the period minimum, and is a bona-fide period bouncer. As a result, it is likely that CVs with donors of spectral type L2 or later are also period bouncers. Indeed, K11 find the donor star's effective temperature at the period minimum is 2140\,K. Using the empirical relationship of \cite{stephens09}, this corresponds to a spectral type of L1; therefore there is a reasonable chance that CVs with donors of spectral type L1 and later have evolved past the orbital period minimum. To our knowledge, there is no other CV with strong {\em spectroscopic} evidence for a donor star of L2 or later. 
A spectral type of M9 has been claimed for VY Aqr \citep{mennickent02}, although this is disputed \citep{harrison09}. Claims for a late-type secondary star in EF Eri were later shown to be due to cyclotron emission \citep{campbell08}. The infrared spectrum of SDSS1212 is also contaminated by cyclotron emission, but there is some evidence that the system contains a very late-type secondary with a spectral type around L8 \citep{farihi08}. A possible case might be made for SDSSJ123813.73+033933.0, since \cite{aviles10} show the infrared photometry is consistent with a spectral type of L4; spectroscopic follow-up observations of this system are highly desirable. 

\section{Conclusions}
\label{sec:concl}
We present time-resolved $J$-band spectroscopy of the short period cataclysmic variable SDSS J1433. We detect absorption lines from the sub-stellar donor star in this system, which contributes $38\pm5$\,\% to the $J$-band light. From the relative strengths of the absorption lines in the $J$-band, we estimate a spectral type of the donor star to be L$2\pm1$. These data are the first spectroscopic detection of a confirmed sub-stellar donor in a cataclysmic variable. The restriction to a single infrared band suggest the formal uncertainty on the spectral type above is an underestimate; follow up spectroscopic observations covering multiple infrared bands are desirable. 

The spectral type indicates an effective temperature of 2000$\pm$200\,K. This is in excellent agreement with the predicted value from semi-empirical evolutionary tracks for CV donor stars. The mass determination for SDSS J1433 makes it an excellent candidate for a period bounce system. Therefore we suggest that other CVs with donor stars of spectral type L2 or later are candidate period bounce systems. From the literature, we find only SDSSJ123813.73+033933.0 has reasonable evidence for a donor star later than L2.

Using skew mapping, we have been able to derive an estimate for the radial velocity of the donor of ${\rm K_d} = 520\pm60$ km\,s$^{-1}$.  This value is consistent with, though much less precise than, predictions from mass determinations found via photometric fitting of the eclipse light curves. 

\section*{\sc Acknowledgements}  This research has made use of
NASA's Astrophysics Data System Bibliographic Services. Based on observations obtained at the Gemini Observatory, which is operated by the Association of Universities for Research in Astronomy, Inc., under a cooperative agreement  with the NSF on behalf of the Gemini partnership: the National Science Foundation  (United States), the National Research Council (Canada), CONICYT (Chile), the Australian Research Council (Australia), Minist\'{e}rio da Ci\^{e}ncia, Tecnologia e Inova\c{c}\~{a}o   (Brazil) and Ministerio de Ciencia, Tecnolog\'{i}a e Innovaci\'{o}n Productiva (Argentina).

\bibliographystyle{mn2e}
\bibliography{refs}

\end{document}